# Guaranteed epsilon-optimal treatment plans with minimum number of beams for stereotactic body radiation therapy

**Hamed Yarmand, David Craft**

Department of Radiation Oncology
Massachusetts General Hospital and Harvard Medical School, Boston, MA 02114, USA

E-mail: yarmand.hamed@mgh.harvard.edu, dcraft@partners.org

**Short title:**

Epsilon-optimal Treatment Plans with Minimum Number of Beams for SBRT

**Corresponding author:**

Hamed Yarmand

Francis H. Burr Proton Therapy Center

Department of Radiation Oncology, MGH

Harvard Medical School

30 Fruit St., Boston, MA 02114, USA

Tell: +1-617-724-3665

Fax: +1-617-724-0368

# Guaranteed epsilon-optimal treatment plans with minimum number of beams for stereotactic body radiation therapy

## H Yarmand[1] and D Craft

Department of Radiation Oncology
Massachusetts General Hospital and Harvard Medical School
Boston, MA 02114, USA

E-mail: yarmand.hamed@mgh.harvard.edu, dcraft@partners.org

**Abstract**

Stereotactic body radiotherapy (SBRT) is characterized by delivering a high amount of dose in a short period of time. In SBRT the dose is delivered using open fields (e.g., beam's-eye-view) known as "apertures". Mathematical methods can be used for optimizing treatment planning for delivery of sufficient dose to the cancerous cells while keeping the dose to surrounding organs at risk (OARs) minimal. Two important elements of a treatment plan are quality and delivery time. Quality of a plan is measured based on the target coverage and dose to OARs. Delivery time heavily depends on the number of beams used in the plan since the setup times for different beam directions constitute a large portion of the delivery time. Therefore the ideal plan, in which all potential beams can be used simultaneously, will be associated with a long impractical delivery time. We use the dose to OARs in the ideal plan to find the plan with the minimum number of beams which is guaranteed to be epsilon-optimal (i.e., a predetermined maximum deviation from the ideal plan is guaranteed). Since the treatment plan optimization is inherently a multicriteria optimization problem, the planner can navigate the ideal dose distribution Pareto surface and select a plan of desired target coverage versus OARs sparing, and then use the proposed technique to reduce the number of beams while guaranteeing epsilon-optimality. We use mixed integer programming (MIP) for optimization. To reduce the computation time for the resultant MIP, we use two heuristics: a beam elimination scheme and a family of heuristic cuts, known as "neighbor cuts", based on the concept of "adjacent beams". We show the effectiveness of the proposed solution technique on two clinical cases, a liver and a lung case. Based on our technique we propose an algorithm for fast generation of epsilon-optimal plans.

---

[1] Corresponding author (yarmand.hamed@mgh.harvard.edu)

# 1 Introduction

Stereotactic body radiotherapy (SBRT) systems have primarily been developed from the extension of stereotactic localization and localization techniques that have been used historically for intracranial radiosurgery (Liu *et al* 2006). Traditional conformal beam radiotherapy planning approaches have often been used, with 5 to 11 beam's-eye-view conformal or intensity-modulated ports (e.g., Shiu *et al* (2003) and Ryu *et al* (2003)). Beam orientation is often chosen manually, and beam arrangements have been either coplanar or noncoplanar depending on the clinical application.

Recently mathematical techniques have been often exploited to obtain high-quality treatment plans by optimizing the beam intensity and/or beam directions. The beam angle optimization problem (BAO) is known to be highly non-convex with many local minima (Sodertrom and Brahme 1993). Therefore finding the optimal solution to BAO requires a relatively long computation time. As a consequence, many heuristics have been developed to find "good" solutions to BAO (e.g., Liu *et al* (2006), Pooter *et al* (2006), and Oldham *et al* (1998)) (it should be noted that BAO has been more investigated for intensity modulated radiation therapy (IMRT) than SBRT, e.g., see Wang *et al* (2005), Liu *et al* (2006), Bertsimas *et al* (In press), Lee *et al* (2006), Lim *et al* (2008), and Lim and Cao (2012)). Oldham *et al* (1998) use a brain SBRT clinical case to investigate the relative benefit of optimizing beam weights, wedge angles, beam orientations, and tomotherapy in improving the dose distribution over the standard plan in which three noncoplanar fields (one open and two wedged) are used. They conclude that the greatest improvement is achieved when beam orientations are optimized.

Liu *et al* (2006) find the optimal number of beams for SBRT of lung and liver lesions based on dose gradient and normal tissue complication probability (NTCP). Their results indicate that while dose gradient improves with more number of beams, there is no significant improvement in NTCP for more than 9 beams for both lung and liver regardless of the target size.

Pooter *et al* (2006) investigate the benefit of using noncoplanar beams for SBRT of liver tumors. They use a heuristic algorithm for automated beam orientation and weight selection to find the optimal plan with a specific number of beams. They conclude that fully noncoplanar beam setups generated by their algorithm are favorable compared to coplanar setups.

In the current literature on BAO, most of which use heuristics, there is no estimate for the optimality gap. The resultant plans are often compared to the standard plans (to demonstrate improvement) rather than to the true optimal plan. In this paper we follow a novel approach to BAO which not only gives an upper bound for the optimality gap in terms of dose to organs at risk (OARs), but also allows to specify a maximum optimality gap and find the plan with the specified maximum optimality gap and the minimum number of beams.

First we find the ideal plan in which all candidate beams (e.g., 36 coplanar beams) can be used simultaneously. Since the treatment plan optimization is inherently a multicriteria optimization problem, the planner can navigate the ideal dose distribution Pareto surface and select the plan which represents the most desirable compromise between target coverage and organ at risk sparing (see Craft *et al* (2012) for details). We have previously introduced interactive multi-criteria-optimization (MCO) to facilitate radiotherapy planning (see Küfer *et al* (2003) for details). In the second step we minimize the number of beams while ensuring the obtained plan is epsilon-optimal with respect to the dose to OARs (e.g., we ensure that the dose to OARs in the obtained plan does not exceed the dose to OARs in the ideal plan for more than epsilon percent) while retaining the coverage constraints exactly. This user-chosen epsilon-optimality guarantee of the final SBRT plan is the most significant improvement over the existing SBRT optimization algorithms.

We formulate BAO as a mixed integer program (MIP) (see Lee *et al* (2006), Lim *et al* (2008) , and Lim and Cao (2012) for examples of MIP formulation for BAO). Since the resultant MIP for BAO is large-scale, heuristics have often been used to obtain good solutions in a reasonable amount of time. For example, Lim *et al* (2008) follow a beam elimination approach and use a scoring method to iteratively eliminate insignificant angles until a predetermined number of beams is reached. However, they admit that iterative methods do not work well with large number of candidate beams. Lim and Cao (2012) use a branch and prune technique (based on a merit score function) combined with local search in a two-phase approach to find clinically acceptable solutions.

One way to reduce the computation time for a large-scale MIP is introducing constraints known as "valid inequalities". The purpose of adding these inequalities is to shrink the feasible region of the linear program (LP) obtained by relaxing the integrality constraint so that it becomes closer to the convex hull of the feasible (integer) solutions, hence obtaining better bounds in the branch and bound (B&B) tree. This method, often combined with heuristics, has sometimes been used in radiation therapy treatment planning for reducing the computation time. For example, Gozbasi (2010) derives valid inequalities for the MIP formulation of the volumetric-modulated arc therapy (VMAT). Tuncel *et al* (2012) derive valid inequalities for FMO under dose-volume restrictions. Lee *et al* (2003) use disjunctive cuts for BAO along with other computational strategies including a constraint and column generation technique. Finally Taskin *et al* (2010, 2011, 2012) develop valid inequalities for the segmentation problem (i.e., the problem of converting the optimal fluence map to deliverable apertures).

In a working paper (Yarmand and Craft ), we introduced two novel heuristic approaches which could be used alone or combined with other MIP-based BAO algorithms to obtain high-quality treatment plans in a reasonable amount of time. One approach is to add a set of heuristic inequalities, referred to as "neighbor cuts", to the associated MIP. The idea is to exploit the intuition that it is less likely that adjacent beams are simultaneously chosen in the optimal beam orientation. The other approach is a beam elimination scheme in which beams with insignificant (dose) contribution in the ideal plan are eliminated from consideration. Then the MIP is solved for the remaining candidate beams. We use both of these heuristics to reduce the computation time. Since the epsilon-optimality is enforced by adding the corresponding constraints to the MIP, the resultant heuristic solution would also be epsilon-optimal. Our numerical results for the clinical cases we have investigated (a liver case and a lung case) show that both of these approaches in general reduce the computation time considerably. For clinical implementation of our technique, we propose an algorithm for fast generation of epsilon-optimal plans.

## 2   Methods and Materials

In this section we discuss the data we used as well as our modeling framework. The proposed technique to find epsilon-optimal treatment plans with the minimum number of beams, as the main contribution of this paper, is presented in Section 2.4, with two refinements to reduce computation time presented in Sections 2.5 and 2.6.

### 2.1   Input data and dose calculation

The planning target volume (PTV) and OARs were identified using three-dimensional images (computed tomography (CT)). These volumetric data were then discretized into voxels for setting up the model instances which include a liver case and a lung case. We used downsampling to reduce the computational burden. Table 1 represents the list of structures for the two model instances along with the number of voxels considered in each of the structures and the model parameters according to the clinical practice at Massachusetts General Hospital.

For the numerical experiments uniformly distributed coplanar beams were considered. For the liver case, there were 34 beams with a gap of 83.72° between the 6[th] and 7[th] beams and a distance of 8.37° between two consecutive beams. For the lung case, there were 24 beams with a gap of 130° between the 24[th] and 1[st] beams and a distance of 10° between two consecutive beams. We emphasize that the methodology and solution technique presented in this paper can be directly applied to noncoplanar beams as well. Each beam was divided into 82-144 beamlets of size $1 \times 1\ cm^2$ (all beamlets go through PTV). For each beamlet, we used CERR (A Computational Environment for Radiotherapy Research) (Deasy *et al* 2003) to calculate the dose per monitor unit intensity to each voxel. An *aperture* is a set of beamlets of a beam which are located consecutively in each row. For each aperture, the dose per monitor unit intensity to a voxel is equal to the sum of the dose per monitor unit intensity from all beamlets in that aperture to that voxel. The total dose per intensity deposited to a voxel is equal to the sum of dose per intensity deposited from each aperture.

**Table 1**. Discretization of anatomical structures and model parameters.

| Case 1: Liver | Number of voxels | [Min,Max] dose (Gy) | Weight in Obj. Fun. | Case 2: Lung | Number of voxels | [Min,Max] dose (Gy) | Weight in Obj. Fun. |
|---|---|---|---|---|---|---|---|
| PTV | 12713 | [50,60] | --- | PTV | 3254 | [70,90] | --- |
| Body | 990 | [0,60] | --- | Shell | 1685 | [0,90] | 0.7 |
| Chest wall | 1893 | [0,60] | 0.99 | Cord | 413 | [0,45] | 0.0 |
| Cord | 335 | [0,45] | 0.0 | Left lung | 3032 | [0,90] | 0.15 |
| Liver | 743 | [0,60] | 0.01 | Chest wall | 1060 | [0,90] | 0.15 |

## 2.2 Generating the pool of apertures

In SBRT open fields or apertures that block a portion of the field are used to deliver the dose. Each aperture can be considered as a set of beamlets in a beam. Beamlets can be on/off in each beam to create apertures with different shapes. The only limitation is that in each row (corresponding to one leaf pair of the multileaf collimator) beamlets which are on should be located consecutively. To generate the pool of diverse candidate apertures first we calculated a *contribution score* for each beamlet based on its contribution to the dose delivery to PTV and OARs. Let $N$ and $L$ denote the set of candidate beams and the set of OARs, respectively. For each beam $i \in N$, let $\bar{A}_i$ denote the set of beamlets associated with beam $i$ and $\bar{D}_{ijk}$ denote the dose per monitor unit intensity contribution to voxel $k$ from beamlet $j$ in beam $i$. The contribution score of beamlet $j$ in beam $i$, denoted by $\alpha_{ij}$, is calculated as follows.

$$\alpha_{ij} = \sum_{k \in V_p} \bar{D}_{ijk} - \sum_{l \in L} \left( \lambda_l \sum_{k \in V_l} \bar{D}_{ijk} \right) \quad , \quad i \in N, j \in \bar{A}_i, \qquad (1)$$

where $V_p$ and $V_l$ denote the set of voxels in PTV and in structure $l \in L$, respectively, and $\lambda_l \geq 0$ denotes the weight of structure $l \in L$ in calculating the contribution score. In the next step we employed a maximum subsequence sum algorithm to find the subsequence of beamlets in each row with the maximum subsequence sum of contribution scores. These subsequences formed the aperture with the maximum sum of beamlets' contribution scores.

As equation (1) demonstrates, the contribution scores, and hence the resultant pool of apertures, depend on $\lambda_l, l \in L$. If $\lambda_l > 0$, then the beamlets which deliver more dose to structure $l$ would have a lower contribution score, and hence would be less likely to be among the beamlets forming the resultant aperture. In other words, if $\lambda_l > 0$, then the resultant aperture avoids delivering a high level of dose to structure $l$. Accordingly we considered $|L| + 2$ different apertures for each beam: one aperture obtained by setting $\lambda_l = \bar{\lambda} > 0, l \in L$ (the smallest aperture which avoids all OARs), one aperture obtained by setting $\lambda_l = 0, l \in L$ (the largest aperture which is the beam's-eye-view of the target), and $|L|$ apertures obtained by setting $\lambda_{l'} = \bar{\lambda} > 0$ for $l' \in L$ and setting $\lambda_l = 0, l \in L, l \neq l'$ (this aperture only avoids structure $l'$). Therefore the pool of apertures contained $(|L| + 2)|N|$ candidate apertures.

Larger values of $\bar{\lambda}$ increase the importance of avoiding OARs, and hence, result in smaller apertures (except for the beam's-eye-view which remains unchanged). The converse is true for smaller values of $\bar{\lambda}$. In order to find the best value, we considered different values for $\bar{\lambda} \in [0.1, 1000]$. To evaluate the performance of the resultant pool of apertures for each value of $\bar{\lambda}$, we used the objective function as well as the dose distribution associated with the ideal plan in which all $|N|$ beams can be used (see Section 2.3). We found that for both liver and lung cases $\bar{\lambda} = 1$ resulted in the best treatment plan. Thus we generated the pool of apertures with $\bar{\lambda} = 1$.

## 2.3 Finding the ideal plan

In the ideal plan all candidate beams can be used simultaneously. One way to determine the ideal plan is to navigate the ideal dose distribution Pareto surface and select a plan, as the ideal plan, of desired target coverage versus OARs sparing (see Craft *et al* (2012) for details). Another way to determine the ideal plan is to solve a linear program (LP) as follows.

For each beam $i \in N$, let $A_i$ denote the set of apertures associated with beam $i$. In a treatment plan, each aperture $j$ in each beam $i$ has a specific intensity $x_{ij} \geq 0$. We refer to $\boldsymbol{x} = (x_{ij}), i \in N, j \in A_i$ as the corresponding treatment plan. If the clinically prescribed lower and upper bounds on the received dose at voxel $k$ are denoted by $C_k$ and $U_k$, respectively, then the basic dosimetric constraints can be represented as follows.

$$\sum_{i \in N} \sum_{j \in A_i} D_{ijk} x_{ij} \geq C_k \, , \, \forall k \text{ and } \sum_{i \in N} \sum_{j \in A_i} D_{ijk} x_{ij} \leq U_k \, , \, \forall k, \tag{2}$$

where $D_{ijk}$ denotes the dose per monitor unit intensity contribution to voxel $k$ from aperture $j$ in beam $i$. The lower and upper bounds on the received dose are usually the same for all voxels in a structure. The values we have used for these bounds for different structures are presented in table 1.

After satisfying constraint (2), it is desired to deliver dose to OARs as small as possible. We use average dose to OARs instead of absolute dose to eliminate the impact of different number of voxels in OARs. We calculate the average dose to OARs by taking average of dose over voxels in each OAR. Let $w_l$ denote the importance weight of structure $l \in L$. Without loss of generality we assume $\sum_{l \in L} w_l = 1$. Then the weighted average dose to OARs under treatment plan $\boldsymbol{x}$, denoted by $D_{OAR}(\boldsymbol{x})$, is

$$D_{OAR}(\boldsymbol{x}) = \sum_{l \in L} w_l \left( \sum_{i \in N} \sum_{j \in A_i} D_{ij}^l x_{ij} \right). \tag{3}$$

The importance weights $w_l, l \in L$ can be determined after the planner has navigated through the ideal Pareto surface and has chosen the desired point. For our numerical experiments we have used the values reported in table 1 which were found by evaluating the dose distribution of the ideal plan. Note that we have considered a zero weight for cord because the dose to cord has been limited to 45 Gy as a constraint.

The objective function of the LP to find the ideal plan is

$$\min D_{OAR}(\boldsymbol{x}), \tag{4}$$

with constraints in (2). This LP is solved very fast using existing software for solving LPs, e.g., CPLEX (version 12.3). Denote the ideal plan by $\boldsymbol{x}^* = (x_{ij}^*), i \in N, j \in A_i$.

To guarantee epsilon-optimality, we need to calculate the dose to OARs in the ideal plan. Let $D_l^*$ denote the dose to structure $l \in L$ in the ideal plan. If the ideal plan is found by navigating the ideal dose distribution Pareto surface, the corresponding dose distribution can be used to calculate $D_l^*, l \in L$. Otherwise, if the corresponding LP is used to find the ideal plan, we calculate $D_l^*, l \in L$, as follows,

$$D_l^* = \sum_{i \in N} \sum_{j \in A_i} D_{ij}^l x_{ij}^* \, , \, l \in L, \tag{5}$$

where $D_{ij}^l$ is the average dose per monitor unit intensity contribution to structure $l \in L$ from aperture $j$ in beam $i$, i.e.,

$$D_{ij}^l = \frac{1}{|V_l|} \sum_{k \in V_l} D_{ijk}. \tag{6}$$

*2.4 Mixed integer programming model for beam reduction*

Since the number of beams in the ideal plan is often too large, we develop a MIP model to find the plan with the desired quality and the minimum number of beams. The developed MIP simultaneously optimizes the beam orientation as well as the beam intensity.

First we discuss the constraints of the MIP. One set of constraints of the MIP are dose level constraints in (2). Another restriction concerns epsilon-optimality of the resultant treatment plan with respect to dose to OARs. We assume the user has specified a maximum deviation of $\varepsilon$ from the optimal dose to OARs as follows.

$$\sum_{i \in N} \sum_{j \in A_i} D_{ij}^l x_{ij} \leq (1+\varepsilon) D_l^* \quad , \quad l \in L, \tag{7}$$

Next we define binary variables to track the number of beams used in the treatment plan. Let $y_i$ be a binary variable whose value is 1 if beam $i$ is used and 0 otherwise. We consider the following constraint to ensure that the aperture intensities are zero for unused beams.

$$x_{ij} \leq M_i y_i \quad , \quad i \in N, j \in A_i, \tag{8}$$

where $M_i$ is a large positive constant and can be chosen based on the maximum possible intensity emitted from beam $i$. We have set $M_i = 35$ for all beams $i \in N$ based on the maximum possible beam intensity.

After discussing the constraints, we focus on the objective function, which concerns the delivery time through the number of beams used in the treatment plan. The number of beams directly impacts the delivery time, and hence is preferred to be as small as possible. It is quite possible that more than one treatment plans use the same optimal number of beams. Therefore we consider a second term in the objective function to ensure that among such plans the one which delivers less amount of dose to OARs is chosen as the optimal plan.

Therefore the objective function of the MIP is

$$\min \sum_{i \in N} y_i + \beta D_{OAR}(x), \tag{9}$$

where $\beta$ is a constant sufficiently small so that the second term is only used to distinguish between different plans with the same optimal number of beams. In other words, existence of the second term should not interfere with finding the plan with the minimum number of beams. More precisely we must have $\beta D_{OAR}(x) < 1$, or equivalently $\beta < [D_{OAR}(x)]^{-1}$. But it follows from (7) that

$$D_{OAR}(x) \leq (1+\varepsilon) \max_{l \in L} D_l^*. \tag{10}$$

Therefore it suffices to choose $\beta < [(1+\varepsilon) \max_{l \in L} D_l^*]^{-1}$.

The objective function in (9) and the constraints in (2), (7), and (8) form the MIP as the basis of our mathematical model. This MIP finds the plan with the minimum number of beams which is guaranteed to be epsilon-optimal with respect to dose to OARs.

*2.5 Neighbor cuts*

The number of potential beams is often much larger than the number of beams which will actually be used. For example, assuming coplanar beams with a grid of 10°, there would be 36 potential beams. However, usually a few of them, e.g., 3 to 7 beams, are selected. Therefore often the selected beams are relatively far apart from each other. In other words, it is less likely that adjacent beams be selected simultaneously in the optimal orientation. This intuition was the basis of our proposed heuristic cuts in (Yarmand and Craft ): we add constraints, referred to as neighbor cuts, to the MIP which allow selection of at most one or a few of the beams in every set of adjacent beams (SAB), i.e., a set whose beams are pairwise adjacent. The general form of neighbor cuts is as follows.

$$\sum_{i \in \Omega_r} y_i \leq T_r \quad , \quad r = 1,2,\dots,R, \tag{11}$$

where $\Omega_r, r = 1,2 \ldots, R$ represent $R$ different SABs and $T_r$ denotes the maximum number of adjacent beams which can be selected from SAB $r$ (note that $T_r < |\Omega_r|$ otherwise constraints (11) will be redundant). The definition of *adjacency* is based on the distance (in degrees) between different beam angles. For example, one might define two adjacent beams as two beams with a distance of at most 10°. In case of uniform distribution of coplanar candidate beams, which is often considered in clinical practice, the adjacency can be defined based on the order in which beams are located around the isocenter. For example, if beams are numbered from 1 to 36, two beams might be defined to be adjacent if their order differs at most by two (it means that every three consecutive beams, e.g., beams 7, 8, and 9, form a SAB). In this case all SABs include the same number of candidate beams, and hence, the same maximum allowed number of beams can be considered for all SABs (i.e., $T_r = T, r = 1,2, \ldots, R$). Therefore in case of uniform and coplanar distribution of candidate beams the neighbor cuts can be represented as follows.

$$\sum_{s=0}^{S-1} y_{i+s} \leq T \ , \ i \in N, \tag{12}$$

where $S \geq 2$ denotes the *neighbor size* defined as the maximum difference in the order of two adjacent beams. Since the beams are repeated after the $N$th beam, we define the index $i + s \equiv i + s - |N|$ if $i + s > |N|$. As discussed before, we have considered coplanar beams for the clinical cases we investigate. However, the definition of adjacent beams, and hence the application of neighbor cuts, can be easily extended to non-coplanar beams in the three-dimensional space.

Two parameters should be determined in order to use the neighbor cuts. One is the neighbor size, $S$, and the other is the maximum allowed number of beams in each SAB, $T$. A larger value for $S$ or a smaller value for $T$ imposes tighter constraints on the original MIP as a result of adding constraints (12). Therefore, in general, it reduces the quality of the heuristic solution and also reduces the computation time further. The converse is in general true for a smaller value for $S$ or a larger value for $T$. It is desired to use the value of $S$ which is large enough to reduce the computation time considerably but small enough to avoid infeasibility.

## 2.6   Beam elimination

In our working paper (Yarmand and Craft ), we also introduced a beam elimination heuristic based on the contribution of different beams to dose delivery in the ideal plan. In the ideal plan, some beams contribute more and some contribute less to delivery of radiation to the tumor. It is quite likely that beams with a lower contribution will not be used if the maximum number of beams is reduced. This was the basis of our beam elimination scheme. After finding the ideal plan, which is found very fast by solving the associated LP, we eliminate the beams with lower contributions to dose delivery. Interestingly enough, only a subset of beams are used in the ideal plan. Therefore we simply eliminate the beams which are not used in the ideal plan and solve the BAO for the remaining beams. Similar to neighbor cuts, the beam elimination can be applied to both coplanar and noncoplanar beams.

## 2.7   Computation

To solve the MIP we used CPLEX (version 12.3) on a PC running Linux with 4 Intel Xeon E5410 (2.33 GHz) CPU and 32 GB of RAM.

## 3 Results and Discussion

The results for the liver and lung cases are presented in Tables 2 and 3, respectively. Rather than computing and navigating an entire Pareto surface, for this study we generate a single Pareto optimal plan, with the number of beams and apertures unrestricted, and call this the ideal plan. These plans for the liver and lung cases are represented in the first row in Tables 2 and 3, respectively. Then we found epsilon-optimal plans with the minimum number of beams for different values of $\varepsilon$ by solving the developed MIP (bold rows in Tables 2 and 3). Finally we incorporated the neighbor cuts and the beam elimination heuristics into the MIP and resolved it. For the neighbor cuts, we considered the gap between the 6$^{th}$ and 7$^{th}$ beams for the liver case and between the 24$^{th}$ and 1$^{st}$ beams for the lung case. We used $\beta = 0.001$ for both liver and lung cases (maximum value for $\beta$ for the liver and lung cases were 0.042 and 0.023, respectively). In Tables 2 and 3, "optimality gap" represents the gap in the objective function of the corresponding solution and the ideal plan. Also "time reduction" represents the reduction in computation time compared to the corresponding MIP with no heuristics (i.e., compared to the corresponding bold rows). Also the underlined and double underlined beams represent beams with two and three apertures in the optimal solution.

**Table 2**. Numerical results for the liver case.

| $\varepsilon$ (%) | S | T | Num. Beams | Num. Aper. | Obj. value | Opt. gap (%) | Comp. time (m) | Time Reduc. (%) | Beams used |
|---|---|---|---|---|---|---|---|---|---|
| 0 | --- | --- | 13 | 19 | 14.21 | 0.00 | 0.63 | --- | 1,2,3,5,6,7,12,13,23,26,27,33,34 |
| **1** | --- | --- | **7** | **13** | **14.26** | **0.38** | **14.84** | --- | **1,3,6,7,13,26,33** |
| 1 | 2 | 1 | 7 | 13 | 14.26 | 0.38 | 3.32 | 77.6 | 1,3,6,7,13,26,33 |
| 1 | 3 | 1 | Infeasible | | | | 1.22 | 91.8 | --- |
| 1 | 3 | 2 | 7 | 13 | 14.26 | 0.38 | 4.22 | 71.6 | 1,3,6,7,13,26,33 |
| 1 | 4 | 2 | 7 | 13 | 14.26 | 0.38 | 3.34 | 77.5 | 1,3,6,7,13,26,33 |
| 1[a] | --- | --- | 7 | 13 | 14.26 | 0.38 | 1.45 | 90.2 | 1,3,6,7,13,26,33 |
| 1[a] | 2 | 1 | 7 | 13 | 14.26 | 0.38 | 1.18 | 92.0 | 1,3,6,7,13,26,33 |
| 1[a] | 3 | 2 | 7 | 13 | 14.26 | 0.38 | 1.22 | 91.8 | 1,3,6,7,13,26,33 |
| 1[a] | 4 | 2 | 7 | 13 | 14.26 | 0.38 | 0.98 | 93.4 | 1,3,6,7,13,26,33 |
| **5** | --- | --- | **4** | **11** | **14.80** | **4.14** | **41.88** | --- | **1,5,7,13** |
| 5 | 2 | 1 | 4 | 11 | 14.80 | 4.14 | 25.49 | 39.1 | 1,5,7,13 |
| 5 | 3 | 1 | 4 | 11 | 14.80 | 4.14 | 9.77 | 76.7 | 1,5,7,13 |
| 5 | 3 | 2 | 4 | 11 | 14.80 | 4.14 | 31.66 | 24.4 | 1,5,7,13 |
| 5 | 4 | 1 | 4 | 11 | 14.80 | 4.14 | 6.76 | 83.9 | 1,5,7,13 |
| 5 | 4 | 2 | 4 | 11 | 14.80 | 4.14 | 27.12 | 35.3 | 1,5,7,13 |
| 5[a] | --- | --- | 4 | 11 | 14.80 | 4.14 | 2.43 | 94.2 | 1,5,7,13 |
| 5[a] | 2 | 1 | 4 | 11 | 14.80 | 4.14 | 2.32 | 94.5 | 1,5,7,13 |
| 5[a] | 3 | 1 | 4 | 11 | 14.80 | 4.14 | 1.73 | 95.9 | 1,5,7,13 |
| 5[a] | 3 | 2 | 4 | 11 | 14.80 | 4.14 | 3.35 | 92.0 | 1,5,7,13 |

[a] Beam elimination has been applied.

Table 3: Numerical results for the lung case.

| ε (%) | (S,T) | T | Num. Beams | Num. Aper. | Obj. value | Opt. gap (%) | Comp. time (m) | Time Reduc. (%) | Beams used |
|---|---|---|---|---|---|---|---|---|---|
| 0 | --- | --- | 7 | 8 | 28.62 | 0.00 | 0.09 | 0.0 | 1,12,13,14,15,18,24 |
| **1** | **---** | **---** | **5** | **6** | **28.62** | **0.02** | **0.20** | **0.0** | **1,12,15,18,24** |
| 1 | 2 | 1 | 5 | 6 | 28.62 | 0.02 | 0.19 | 5.0 | 1,12,15,18,24 |
| 1 | 3 | 1 | 5 | 6 | 28.62 | 0.02 | 0.19 | 7.7 | 1,12,15,18,24 |
| 1 | 3 | 2 | 5 | 6 | 28.62 | 0.02 | 0.19 | 6.4 | 1,12,15,18,24 |
| 1 | 4 | 1 | Infeasible | | | | 0.08 | 62.9 | |
| 1 | 4 | 2 | 5 | 6 | 28.62 | 0.02 | 0.19 | 7.0 | 1,12,15,18,24 |
| 1 | 5 | 2 | 5 | 6 | 28.62 | 0.02 | 0.12 | 39.0 | 1,12,15,18,24 |
| 1[a] | --- | --- | 5 | 6 | 28.62 | 0.02 | 0.04 | 79.6 | 1,12,15,18,24 |
| 1[a] | 2 | 1 | 5 | 6 | 28.62 | 0.02 | 0.03 | 83.1 | 1,12,15,18,24 |
| 1[a] | 3 | 1 | 5 | 6 | 28.62 | 0.02 | 0.03 | 84.1 | 1,12,15,18,24 |
| 1[a] | 3 | 2 | 5 | 6 | 28.62 | 0.02 | 0.03 | 83.7 | 1,12,15,18,24 |
| **5** | **---** | **---** | **3** | **5** | **29.44** | **2.86** | **1.52** | **0.0** | **6,14,18** |
| 5 | 2 | 1 | 3 | 5 | 29.44 | 2.86 | 5.41 | -256.4 | 6,14,18 |
| 5 | 3 | 1 | 3 | 5 | 29.44 | 2.86 | 3.51 | -131.0 | 6,14,18 |
| 5 | 4 | 1 | 3 | 5 | 29.44 | 2.86 | 1.79 | -18.0 | 6,14,18 |
| 5 | 5 | 1 | 3 | 4 | 29.49 | 3.04 | 1.09 | 28.3 | 1,15,24 |
| 5 | 6 | 1 | 3 | 4 | 29.49 | 3.04 | 0.95 | 37.2 | 1,15,24 |
| 5[a] | --- | --- | 3 | 4 | 29.49 | 3.04 | 0.15 | 90.2 | 1,15,24 |
| 5[a] | 2 | 1 | 3 | 4 | 29.49 | 3.04 | 0.18 | 88.4 | 1,15,24 |
| 5[a] | 3 | 1 | 3 | 4 | 29.49 | 3.04 | 0.14 | 90.9 | 1,15,24 |
| 5[a] | 3 | 2 | 3 | 4 | 29.49 | 3.04 | 0.12 | 92.3 | 1,15,24 |

[a] Beam elimination has been applied.

As Tables 2 and 3 show, epsilon-optimal plans use considerably smaller number of beams compared to the ideal plan. As expected, the number of beams is smaller for $\varepsilon = 5\%$ compared to $\varepsilon = 1\%$ due to the larger allowed deviation from the ideal plan. In general, the most reduction in computation time is achieved when the neighbor cuts are combined with the beam elimination. This combination performs very well for both the liver and lung cases. The beam elimination heuristic, when applied alone, has a very good performance in reducing the computation time for both liver and lung cases. The heuristic neighbor cuts, when applied alone, have a good performance for the liver case. However, the neighbor cuts do not perform well for the lung case, especially for $\varepsilon = 5\%$. The reason is that the number of active beams in the ideal plan of the lung case (and also in the optimal plan for $\varepsilon = 1\%$ and $\varepsilon = 5\%$) is relatively small. Therefore feasible solutions, and hence tight upper bounds, are found relatively fast in the B&B tree. As a result, the MIP for $\varepsilon = 1\%$ and $\varepsilon = 5\%$ is solved relatively fast in absence of any heuristics (compare the computation times for the liver and lung cases). Now the beam elimination heuristic reduces the computation time of the MIP further as it reduces the number of potential beams.

However, the neighbor cuts have a dual effect on the computation time of the MIP. They have a favorable effect as they reduce the feasible region of the corresponding relaxed LP resulting in tighter bounds in B&B. Note that this effect is magnified for smaller $\varepsilon$ in which case it is harder to find tight bounds with no heuristics. But the neighbor cuts also have an unfavorable effect as they increase the number of constraints, which results in an increase in the solution time of the corresponding relaxed LP at every node of the B&B tree. If the unfavorable effect is larger than the favorable effect, then incorporating the neighbor cuts into the MIP will increase the computation time. In the lung case with relatively small number of active beams tight bounds are found relatively fast without the neighbor cuts especially for $\varepsilon = 5\%$. Therefore the favorable effect of the neighbor cuts is overshadowed by the unfavorable effect resulting in an increase in the computation time. This observation suggests that neighbor cuts should be used with caution when the number of active beams in the ideal plan is small.

Note that infeasibility as a result of adding the neighbor cuts is determined very fast. Therefore the values of $S$ and $T$ can be modified to find a feasible epsilon-optimal treatment plan.

To illustrate the quality of the resultant epsilon-optimal plans, we have compared the dose-volume histogram (DVH) for the ideal plan of the liver case (which uses 13 beams) with the generated plan with $\varepsilon = 5\%$ (which uses 4 beams) in Figure 1. Note that although the cord dose has gone up significantly, it is still well below the constraint of 45 Gy.

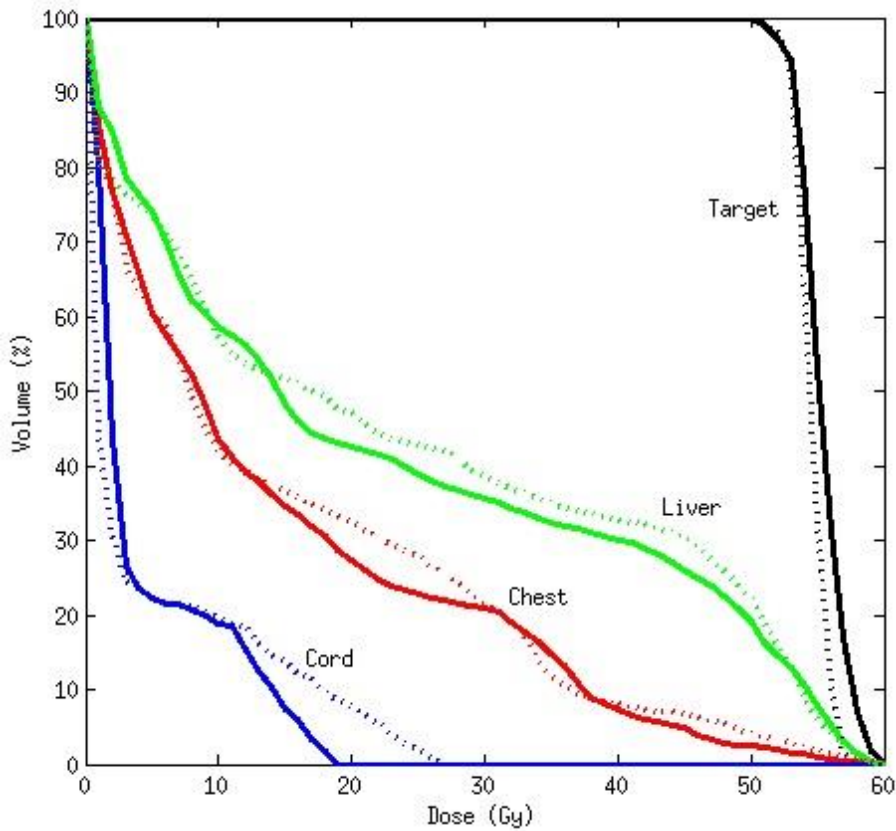

**Figure 1.** DVH for the ideal solution of the liver cases with 13 beams (solid lines) and the epsilon-optimal plan at $\varepsilon = 5\%$ with 4 beams (dotted lines).

## 4 An algorithm for fast generation of epsilon-optimal plans

There are several parameters which should be evaluated for implementing the proposed technique. In clinical practice, after achieving an acceptable target coverage, there are mainly three objectives: (i) low irradiation of OARs (controlled by $\varepsilon$), (ii) short delivery time (controlled by the number of beams), and (iii) short computation time (controlled by $S$ and $T$). Accordingly we have developed an iterative algorithm for fast generation of epsilon-optimal plans as follows.

**Algorithm 1:**

Step 1: Navigate the ideal dose distribution Pareto surface and select a plan of desired target coverage versus OARs sparing (see Craft *et al* (2012) for details). Use the corresponding importance weights, $w_l$, $l \in L$.

Step 2: Find the ideal plan. If the number of beams in the ideal plan is sufficiently small, stop. Otherwise set $\varepsilon \leftarrow \varepsilon_{min}$, $S \leftarrow S_{max}$, $T \leftarrow T_{min}$.

Step 3: Solve MIP.

Step 4: If MIP is feasible and the number of beams in the generated plan is sufficiently small, stop. If MIP is infeasible, set $S \leftarrow S - 1$ or $T \leftarrow T + 1$ and got to Step 3. If MIP is feasible but the number of beams is not sufficiently small, increase $\varepsilon$ and set $S \leftarrow S_{max}$, $T \leftarrow T_{min}$ and go to Step 3.

Note that beam elimination can be applied in Step 2 instead of neighbor cuts or combined with them. As an example of application of Algorithm 1, if for the liver case we set $\varepsilon \leftarrow 1\%$, $S \leftarrow 3$, $T \leftarrow 1$, the MIP will be infeasible as shown in table 2. Then by setting $S \leftarrow 2$, the MIP will be feasible and use 7 beams. Next if we want to decrease the number of beams further, we can set $\varepsilon \leftarrow 5\%$ and reset $S \leftarrow 3$, $T \leftarrow 1$. The resultant MIP will be feasible and use 4 beams as shown in table 2. The DVH of this plan is compared with the DVH of the ideal plan in Figure 1. The total computation time (after the importance weights have been determined) with using the neighbor cuts will be $0.63 + 1.22 + 3.32 + 9.77 = 14.94$ minutes, which is 74% less than the total computation time without using the neighbor cuts ($0.63 + 14.84 + 41.88 = 57.35$ minutes). The generated plan irradiates OARs only 4.14% more than the ideal plan but uses only 4 beams versus 13 beams used in the ideal plan, hence a great saving in setup time.

## 5 Conclusion and future research directions

The proposed technique provides a tool for efficiently generating quality-guaranteed treatment plans for SBRT. This technique can be combined with a general algorithm for navigation of the ideal dose distribution Pareto surface to find a treatment plan with the minimum number of beams and a pre-specified maximum deviation from the desired ideal plan. The beam elimination and neighbor cuts heuristics in general reduce the computation time considerably while preserving the quality guarantee. As a result, quality-guaranteed treatment plans with the minimum delivery time can be obtained very fast by applying Algorithm 1. The beam elimination heuristic can also be applied to nonlinear (e.g., quadratic) models.

The proposed technique can also be applied to IMRT treatment planning with apertures replaced with beamlets to find epsilon-optimal treatment plans. Regarding the heuristics, neighbor cuts can be applied as they were applied for SBRT. However, it would be more complicated to apply the beam elimination heuristic because in the IMRT ideal plan all beams will be used. Therefore the beam elimination should be applied based on the contribution of each beam in delivering the dose to the tumor. In such a circumstance, the efficiency of the beam elimination heuristic might be reduced. Evaluation of efficiency of the proposed technique, and in particular the heuristics, for IMRT is an interesting direction for future research.

**Acknowledgments**

The project was supported by the Federal Share of program income earned by Massachusetts General Hospital on C06 CA059267, Proton Therapy Research and Treatment Center and partially by RaySearch Laboratories.

**References**

Bertsimas, D, Cacchiani, V, Craft, D and Nohadani, O In press A hybrid approach to beam angle optimization in intensity-modulated radiation therapy *Computers & Operations Research*

Craft, D, McQuaid, D, Wala, J, Chen, W, Salari, E and Bortfeld, T 2012 Multicriteria VMAT optimization *Med. Phys.* **39** 2 686

Deasy, J O, Blanco, A I and Clark, V H 2003 CERR: A computational environment for radiotherapy research *Med. Phys.* **30** 5 979

Gozbasi, H O 2010 Optimization approaches for planning external beam radiotherapy *Doctor of Philosophy* Georgia Institute of Technology

Küfer, A H, Scherrer, A, Monz, M, Alonso, F, Trinkaus, H, Bortfeld, T and Thieke, C 2003 Intensity-modulated radiotherapy - a large scale multi-criteria programming problem *OR Spectrum* **25** 2 223-49

Lee, E K, Fox, T and Crocker, I 2006 Simultaneous beam geometry and intensity map optimization in intensity-modulated radiation therapy *International Journal of Radiation Oncology Biology Physics* **64** 1 301-20

Lee, E K, Fox, T and Crocker, I 2003 Integer Programming Applied to Intensity-Modulated Radiation Therapy Treatment Planning *Annals of Operations Research* **119** 165-81

Lim, G J and Cao, W 2012 A two-phase method for selecting IMRT treatment beam angles: Branch-and-Prune and local neighborhood search *European Journal of Operational Research* **217** 609-18

Lim, G J, Choi, J and Mohan, R 2008 Iterative solution methods for beam angle and fluence map optimization in intensity modulated radiation therapy planning *OR Spectrum* **30** 289-30

Liu, H, Jauregui, M, Zhang, X, Wang, X, Dong, L and Mohan, R 2006 Beam angle optimization and reduction for intensity-modulated radiation therapy of non-small-cell lung cancers *International Journal of Radiation Oncology Biology Physics* **65** 2 561-72

Liu, R, Buatti, J M, Howes, T L, Dill, J, Modrick, J M and Meeks, S L 2006 Optimal number of beams for stereotactic body radiotherapy of lung and liver lesions *International Journal of Radiation Oncology Biology Physics* **66** 3 906-12

Oldham, M, Khoo, V S, Rowbottom, C G, Bedford, J L and Webb, S 1998 A case study comparing the relative benefit of optimizing beam weights, wedge angles, beam<br />orientations and tomotherapy in stereotactic radiotherapy of the brain *Physics in Medicine and Biology* **43** 2123-46

Pooter, J A, Romero, A M, Jansen, W P A, Storchi, P R M, Woudstra, E, Levendag, P C and Heijmen, B J M 2006 Computer optimization of noncoplanar beam setups improves stereotactic treatment of liver tumors *International Journal of Radiation Oncology Biology Physics* **66** 3 913-22

Ryu, S, Fang Yin, F, Rock, J, Zhu, J, Chu, A, Kagan, E, Rogers, L, Ajlouni, M, Rosenblum, M and Kim, J H 2003 Image-guided and intensity-modulated radiosurgery for patients with spinal metastasis *Cancer* **97** 8 2013-8

Shiu, A S *et al* 2003 Near simultaneous computed tomography image-guided stereotactic spinal radiotherapy: An emerging paradigm for achieving true stereotaxy *International Journal of Radiation Oncology\*Biology\*Physics* **57** 3 605-13

Sodertrom, S and Brahme, A 1993 Optimization of the dose delivery in a few field techniques using radiobiological objective functions *Med. Phys.* **20** 4 1201-10

Taskin, Z C, Smith, J C, Romeijn, H E and Dempsey, J F 2010 Optimal Multileaf Collimator Leaf Sequencing in IMRT Treatment Planning *Oper. Res.* **58** 3 674-90

Taskin, Z C and Cevik, M 2011 Combinatorial Benders cuts for decomposing IMRT fluence maps using rectangular apertures *Comput. Oper. Res.* **In press**

Taskin, Z C, Smith, J C and Romeijn, H E 2012 Mixed-integer programming techniques for decomposing IMRT fluence maps using rectangular apertures *Annals of Operations Research* **196** 1 799-818

Tuncel, A T, Preciado, F, Rardin, R L, Langer, M and Richard, J P 2012 Strong valid inequalities for fluence map optimization problem under dose-volume restrictions *Annals of Operations Research* **196** 1 819-40

Wang, X, Zhang, X, Dong, L, Liu, H, Gillin, M, Ahmad, A, Ang, K and Mohan, R 2005 Effectiveness of noncoplanar IMRT planning using a parallelized multiresolution beam angle optimization method for paranasal sinus carcinoma *International Journal of Radiation Oncology Biology Physics* **63** 2 594-601

Yarmand, H and Craft, D Two Effective Heuristics for Beam Angle Optimization in Radiation Therapy *Working paper*